\documentclass[a4paper,11pt]{article}
\usepackage{pos}
\usepackage{natbib}
\usepackage{mathtools}
\usepackage{floatrow}
\usepackage{wrapfig}
\usepackage{booktabs}
\usepackage{abbrv}

\title{Cosmic ray calorimetry in star-forming galaxy populations and implications for their contribution to the extra-galactic $\gamma$-ray background}
 \ShortTitle{The extra-galactic $\gamma$-ray background: effects of CR calorimetry in SFGs}

\author*[a]{Ellis R. Owen}
\author[b,c]{Albert K. H. Kong}
\author[b,c,d]{Kuo-Chuan Pan}

\affiliation[a]{Theoretical Astrophysics, Department of Earth and Space Science, Graduate School of Science, Osaka University, Toyonaka,
Osaka 560-0043, Japan}

\affiliation[b]{Institute of Astronomy, National Tsing Hua University, Hsinchu 30013, Taiwan (ROC)}

\affiliation[c]{Center for Informatics and Computation in Astronomy, \\ National Tsing Hua University, Hsinchu 30013, Taiwan (ROC)}

\affiliation[d]{Physics Division, National Center for Theoretical Sciences, Taipei 10617, Taiwan (ROC)}

\emailAdd{erowen@astro-osaka.jp}

\abstract{Star-forming galaxies (SFGs) have been established as an important source population in the extra-galactic $\gamma$-ray background (EGB). Their intensive star-formation creates an abundance of environments able to accelerate particles, and these build-up a rich sea of cosmic rays (CRs). Above GeV energies, CR protons can undergo hadronic interactions with their environment to produce $\gamma$-rays. SFGs can operate as CR proton "calorimeters", where a large fraction of the CR energy is converted to $\gamma$-rays. However, CRs also deposit energy and momentum to modify the thermal and hydrodynamic conditions of the gas in SFGs, and can become a powerful driver of outflows. Such outflows are ubiquitous among some types of SFGs, and have the potential to severely degrade their CR proton calorimetry. This diminishes their contribution to the EGB. In this work, we adopt a self-consistent treatment of particle transport in outflows from SFGs to assess their calorimetry. We use 1D numerical treatments of galactic outflows driven by CRs and thermal gas pressure, accounting for the dynamical effects and interactions of CRs. We show the impact CR-driven flows have on the relative contribution of SFG populations to the EGB, and investigate the properties of SFGs that contribute most strongly.}

\ConferenceLogo{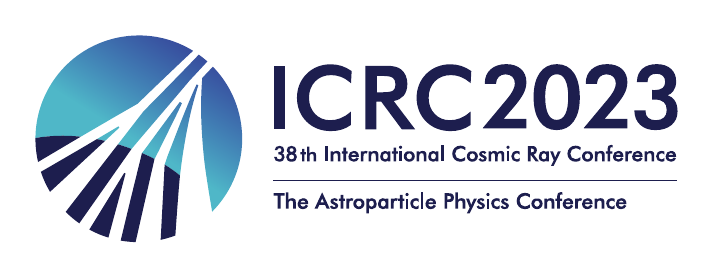}

\FullConference{%
38th International Cosmic Ray Conference (ICRC2023)\\
  26 July - 3 August, 2023\\
  Nagoya, Japan}

\begin{document}
\maketitle

\section{Introduction}
\label{sec:introduction}

\vspace{-0.2cm}
\noindent
Cosmic rays (CRs) are an important ingredient in the galaxy evolution recipe. They modify the thermal and hydrodynamical properties of the interstellar and circum-galactic media of galaxies~\cite[e.g.][]{Owen2018MNRAS}, and produce photonic and non-photonic signatures. 
Astrophysical environments rich in interacting CRs emit $\gamma$-rays. 
Star-forming galaxies (SFGs) are an example of this, and several nearby SFGs 
have been resolved in $\gamma$-rays with \textit{Fermi}-LAT~\cite{Ajello2020ApJ_SFG}. CRs in SFGs are likely accelerated through diffusive shock acceleration processes in sources like supernovae (SNe), SN remnants, and young massive stellar clusters.

The accumulated unresolved emission of $\gamma$-rays from SFG populations
has been considered to make a non-negligible contribution to the extra-galactic $\gamma$-ray background (EGB), possibly exceeding several tens of percent between 1-100 GeV~\cite{Owen2021MNRAS, Owen2022MNRAS, Roth2021Natur, Sudoh2018PASJ, Peretti2020MNRAS}. Besides SFGs, other sources such as Seyferts and blazars also contribute to the EGB~\citep[e.g.][]{Lamastra2017A&A}, with many such sources having been resolved. However, the exact balance of source types forming the remaining unresolved EGB is yet to be firmly established. 
New data will soon be obtained with upcoming instruments like the Cherenkov Telescope Array (CTA). It is therefore important to develop models to assess the possible contributions from a range of source populations. This will ensure data can be properly interpreted, and appropriate tools are developed to efficiently extract meaningful information about the origins of the EGB. As a potentially dominant EGB source populations~\cite[e.g.][]{Roth2021Natur}, SFGs are particularly worthy of consideration. 

In previous work, the contribution of SFGs to the EGB was modeled using a prototype approach~\cite[see, e.g.][]{Peretti2020MNRAS, Ambrosone2021MNRAS, Roth2021Natur}. Key parameters (e.g. star-formation rate, stellar mass, injected CR spectral index) were used to specify the 
$\gamma$-ray emission of galaxies, with some studies 
also taking into account 
intervening dependencies on 
effective galaxy size, molecular gas density and
dust properties~\cite[see][]{Owen2021MNRAS, Owen2022MNRAS}. 
In many of these previous models, a tuning parameter was applied uniformly to represent the confinement and calorimetry of CRs. Setting this value below unity would represent the fractional loss of CRs from a galaxy by advection and/or diffusive leaking before they undergo a hadronic interaction. However, some galaxies can approach a calorimetric limit where CRs are almost entirely absorbed before escaping~\cite{Krumholz2020MNRAS}. This leads to a very high fraction a galaxy's CR luminosity being converted to $\gamma$-rays. 
In this study, we invoke a prototype treatment that self-consistently accounts for CR calorimetry to model the SFG contribution to the EGB. This is an extension of previous work~\cite[Ref.][]{Owen2022MNRAS}, where earlier fixed assumptions about CR escape in galactic outflows have now been relaxed. 

\section{Prototype model}
\label{sec:prototype_model}

\vspace{-0.2cm}
\noindent 
Our calculations follow the methodology introduced in \cite{Owen2022MNRAS, Owen2021MNRAS}. We refer the reader to these earlier works for a detailed description of our computational implementation, galaxy population models, and our treatment of CR propagation and interactions.  
In SFGs, $\gamma$-rays are produced by two channels: (1) energetic hadronic CRs interacting by proton-proton (pp) pion-production processes with interstellar gas, and (2) leptonic inverse-Compton processes from high energy CR electrons interacting with ambient radiation fields (primarily the interstellar radiation field of the host galaxy, and the cosmological microwave background). Our model includes both these contributions. The density of interstellar gas is estimated from the galaxy's mass and star-formation rate, while the radiation field is specified by the galaxy's redshift, star-formation rate, interstellar dust composition, and the average temperature of its starlight. 
Table~\ref{tab:param} summarizes the fixed model parameters, while other aspects of SFG populations are derived from outputs of cosmological simulations of galaxy formation (see Ref.~\cite{Owen2021MNRAS} for a complete description).
In this section, we describe the aspects of our model that differ from our previous work. 

\begin{table}
\centering
\resizebox{0.7\textwidth}{!}{%
\begin{tabular}{*{3}{|c}|}
\hline 
{\bf Parameter} & {\bf Value} & {\bf Definition} \\
\hline 
$\Gamma$ & $-2.1$ & CR proton spectral index \\
$\gamma_{\rm p}^{\star}$ & $10~\text{PeV}/m_{\rm p}c^2$ & Maximum CR proton energy \\ 
$D$ & $3.0\times 10^{28}$ cm$^2$ s$^{-1}$ & CR diffusion coefficient \\
$r_{\rm sb}$ & 300 pc & Size for the starburst region of a SFG  \\
$f_t$ & 0.1 & Turbulent to magnetic energy transfer efficiency \\
$\alpha$ & {0.05} & Fraction of stars that produce a core-collapse SN event \\
$M_{\rm SN}$ & $50 \;\! {\rm M}_{\odot}$ & Upper mass of stars able to produce a SN event \\
$E_{\rm SN}$ & $10^{53}\;\!{\rm erg}$ & Total energy of a core-collapse SN  \\
$\varepsilon$ & 0.1 & CR acceleration efficiency in SN remnants  \\
$f_{\nu}$ & $0.01$ & SN kinetic energy available after losses to neutrinos \\
{$\kappa_{\rm e}$} & $0.034$ & {Fraction of total CR energy supplied to primary electrons} \\
$f_{\rm abs}$ & 0.26 & Fraction of ionising stellar photons absorbed by H \\
$\beta$ & 0.6 & Average dust absorption efficiency of non-ionising photons \\ 
$\eta$ & 0.5 & Fraction of infra-red emission from diffuse ISM gas \\
$T^{\star}$ & $30,000\;\!{\rm K}$ & Temperature of the stellar radiation field \\
$n_{\rm cl}$ & $10\;{\rm cm}^{-3}$ & Density of cold entrained clumps in outflows~\cite{Melioli2013MNRAS}$^{(a)}$ \\
$f_{\rm c}$ & $0.1$ & Volume filling fraction of cold clumps in outflow~\cite{Martin1998ApJ}$^{(b)}$ \\
\hline
\end{tabular}%
}
\caption{A list of fixed physical parameters adopted in our prototype galaxy model. Other model parameters are allowed to vary. \\
{\footnotesize \textbf{Notes:} ${(a)}$ The adopted value is the upper end of the range indicated by Ref.~\cite{Melioli2013MNRAS} to provide a maximum EGB estimate. 
${(b)}$ This was adopted by Ref.~\cite{Martin1998ApJ} as the clumping factor for H$\alpha$ emitting gas, not the dense clumps. However, in the absence of any physical constraint on this parameter, we consider 
that it is not any more or less physically meaningful to assume a similar volume fraction of dense clumps to H$\alpha$ stripped filaments as an indicative choice compared to other values. Detailed simulation work and charge exchange observations will allow for much improved  constraints in future.}}
\label{tab:param}
\end{table} 

\subsection{Galaxy outflow model}

\vspace{-0.2cm}
\noindent 
The combined effects of thermal and CR pressure arising from the confluence of feedback from the concentrated star-formation in a SFG can drive a large-scale galactic outflow. This transfers energy, matter, magnetic fields and CRs from the interior of a galaxy to its halo, and reduces the calorimetric capability of a galaxy to CRs. 
Galactic outflows can be modeled as a 2-fluid system, comprised of a thermal gas and non-thermal CR fluid. The thermal fluid may be multi-phase, however we do not consider the dynamical effects of an entrained dense phase. In this prescription, the hydrodynamical (HD) equations for thermal gas component are written as:  
\begin{align}
\label{eq:HD_mass}
\frac{\partial\rho}{\partial t}+\nabla\cdot\left(\rho{\boldsymbol v}\right) & =q\ ,\\
\label{eq:HD_momentum}
\frac{\partial\rho{\boldsymbol v}}{\partial t}
  +\nabla\cdot\left(\rho{\boldsymbol v}{\boldsymbol v}\right) & =\rho {\boldsymbol g}-\nabla P_{\rm tot}\ ,\\
\label{eq:HD_energy}
\frac{\partial e_{\rm g}}{\partial t}
 +\nabla\cdot\left[\left(e_{\rm g}+P_{\rm g} \right){\boldsymbol v}\right]
  & =Q-C+\rho{\boldsymbol v}\cdot{\boldsymbol g}+\mathcal{I} \ ,
\end{align}
and the energy equation for the CR fluid is written as: 
\begin{align}
     \label{eq:CR_energy}
  \frac{\partial e_{\rm c}}{\partial t} + \nabla \cdot \left[  \left(e_{\rm c} + P_{\rm c} \right) {\boldsymbol v}_{\rm c} \right] 
  & = \nabla \cdot \left[ D\nabla e_{\rm c}\right] - \mathcal{I} - f_{\rm c} n_{\rm cl} c \sigma_{\rm pp} e_{\rm c} + Q_{\rm c} \ .
\end{align}
Here, it is assumed that (1) CRs diffuse isotropically at a rate set by the diffusion coefficient $D$, (2) that the magnetic field acts only as a coupling mechanism between the CR and thermal fluid, and (3) that CRs are not affected by the gravitational potential of their host galaxy.\footnote{We consider this is reasonable, as their energy will not be significantly affected by work done to move out of the gravitational potential of the galaxy.} 
The term $f_{\rm c} n_{\rm cl} c \sigma_{\rm pp} e_{\rm c}$ accounts for hadronic losses of CRs in the dense entrained phase of the outflow, where $c$ is the speed of light, $\sigma_{\rm pp}$ is the pp interaction cross section, $n_{\rm cl} = 10~ {\rm cm}^{-3}$ is the gas number density in the entrained clouds (estimated from the range found by Ref.~\cite{Melioli2013MNRAS}), and $f_{\rm c} = 0.1$ is the volume filling factor of the dense clouds in the entrained flow (this is the same as Ref.~\cite{Martin1998ApJ}). 
In the above system of equations, 
the total pressure is 
$P_{\rm tot} = P_{\rm g} + P_{\rm c} + B^2/8\pi$, where 
$P_{\rm g} = (\gamma_{\rm g}-1)e_{\rm g}$, and 
$P_{\rm c} = (\gamma_{\rm c} -1) e_{\rm c}$ are the thermal gas and CR pressure, respectively. $e_{\rm g} = 0.5 \rho v^2 + e_{\rm th}$, is the energy density of the thermal gas, $e_{\rm c}$ is the energy density of the CRs, 
  $\rho$ is gas density,
 ${\boldsymbol v}$ is gas velocity, and ${\boldsymbol v}_{\rm c} = {\boldsymbol v} + {\boldsymbol v}_{\rm A}$ is the CR velocity, being the sum of the entraining gas velocity and the local Alfv\'{e}n speed (corresponding to the effective large-scale CR streaming speed), 
 and $C$ is the radiative cooling rate of the thermal gas.  $\mathcal{I} = -({\boldsymbol v} + {\boldsymbol v}_{\rm A})\cdot \nabla P_{\rm c} + \mathcal{C}_{\rm c}  e_{\rm c}$ is the energy transfer rate from the CRs to the thermal component of the wind   
  \citep[see][]{Yu2020MNRAS}. ${\boldsymbol g}$ 
 is  the external (gravitational) force term. The gravitational force is 
  provided by the dark matter halo of the galaxy, 
  which takes the form of a Navarro-Frenk-White profile, where the total halo mass is obtained from the stellar mass by the fitted function provided by Ref.~\cite{Zaritsky2023MNRAS}. This is valid for stellar masses below 10$^9$ M$_{\odot}$, or halo masses below 10$^{12}$ M$_{\odot}$, which covers the range in which the CR calorimetry fraction is found to be below unity. Galaxies with larger masses than this cannot typically launch an outflow by the mechanisms discussed in this work.  

 The source terms are given by
$q(z) = {3 \dot{M} \mathcal{S}(r)}/{\pi {r_{\rm sb}}^3}$, 
$Q(z) = {3 \dot{E} \mathcal{S}(r)}/{\pi {r_{\rm sb}}^3}$
and 
$Q_{\rm c}(z) = {3 \dot{E_{\rm c}} \mathcal{S}(r)}/{\pi {r_{\rm sb}}^3}$ for mass, 
  energy and CR energy respectively, where 
 $\mathcal{S}(r) = \left[1-\mathcal{H}(|r-r_{\rm sb}|)\right]$.  
 $\dot{M}$ is the total mass injection rate, 
  and $\dot{E}$ and $\dot{E}_{\rm c}$ are respectively 
  the energy injection rates into the thermal gas and CR fluid 
  across the entire starburst region. 
 $\mathcal{H}(\cdots)$ is the Heaviside step function, and $r_{\rm sb} = 300$ pc. 
The mass injection rate, $\dot{M}$, and the total energy injection rate, $\dot{E}_{\rm tot} = \dot{E} + \dot{E}_{\rm c}+L_0$, are parametrised by the star formation rate $\mathcal{R}_{\rm SF}$, and are given by 
    $\dot{M} = 0.26 ~ \mathcal{R}_{\rm SF} \; {\rm M}_\odot\ \rm yr^{-1} \ ,$
where 0.26 is the estimated fraction of mass lost to stellar winds and supernovae (computed using {\tt Starburst99}\footnote{\url{https://www.stsci.edu/science/starburst99/}}, with solar metallicity~\cite{Leitherer1999ApJS}), and
    $\dot{E}_{\rm tot}=7\times10^{41}~\left({\mathcal{R}_{\rm SF}}/{{\rm M}_\odot\ \rm yr^{-1}}\right) ~{\rm erg}\;\!{\rm s}^{-1}$
\citep{Leitherer1999ApJS}, with values scaled from M82. 90 per cent of the total injected energy is radiated away \citep{Leitherer1999ApJS}. The remainder is therefore available to drive an outflow. Following the fiducial choices of~\cite{Yu2021MNRAS}, we take the CR and thermal powers to be equal, i.e.
$\dot{E} = \alpha \dot{E}_{\rm tot}$, 
$\dot{E}_{\rm c} = \alpha_{\rm c} {\dot{E}_{\rm tot}}$, and 
$L_{0} = \alpha_{\rm L} {\dot{E}_{\rm tot}}$, 
where $\alpha = 0.05$, $\alpha_{\rm c} = 0.05$, and $\alpha_{\rm L} = 0.9$.

\subsection{Outflow structure and evolution, and cosmic ray calorimetry}

\begin{figure}
    \centering
    \includegraphics[width=\textwidth]{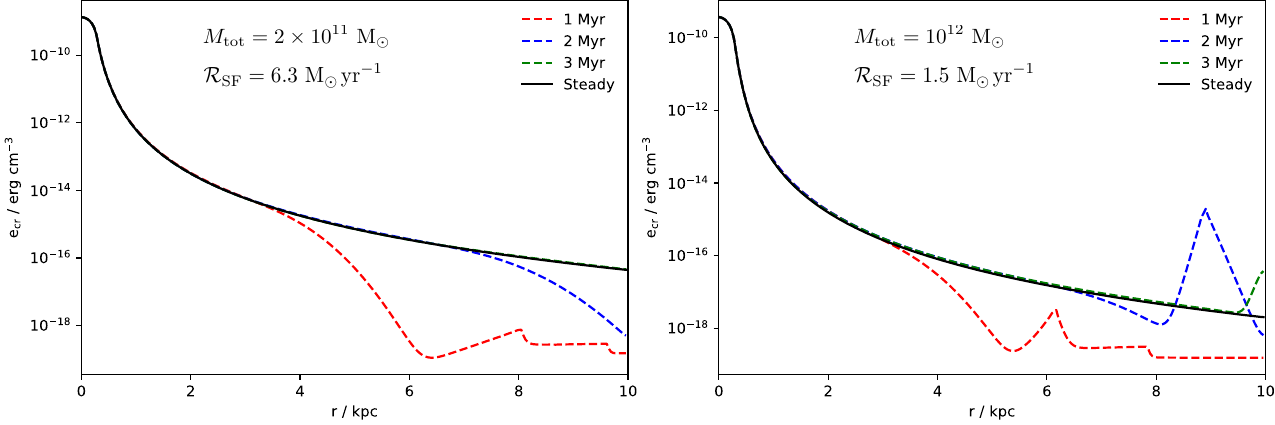} 
        \caption{Outflows evolve to a steady or unsteady state after their initial eruption from the starburst region. The starburst episode driving the outflow is considered to persist for 10s of Myr. The flow evolution and profile is affected by the properties of the host galaxy, modifying the distribution and escape of CRs. The left panel shows a more intensively star-forming galaxy in a weaker gravitational potential. Compared to the right panel, the flow evolves faster and reaches a steady state more quickly.}
  \label{fig:outflow_evo}
\end{figure}

\vspace{-0.2cm}
\noindent
We use {\tt FLASH}~\cite{Fryxell2000ApJS}\footnote{\url{https://flash.rochester.edu/site/}}, a Eulerian grid code, to solve equations~\ref{eq:HD_mass} to~\ref{eq:CR_energy}. We pre-compute outflow solutions across a parameter grid, varying galaxy star-formation rate and mass. 
For some parameter choices, a steady state solution is achieved shortly after an initial outflow eruption phase, generally within a few Myr (see Fig.~\ref{fig:outflow_evo}). Conversely, certain parameter combinations do not yield a stationary outflow solution. In these cases, 
we computed the distribution of CRs 
in the outflow over a period of 100 Myr after the eruption phase. This allowed us to gauge the long-term, time-averaged CR distribution for the non-steady flow. 

\begin{figure}
\floatbox[{\capbeside\thisfloatsetup{capbesideposition={left,bottom},capbesidewidth=6.5cm}}]{figure}[\FBwidth]
{\caption{Calorimetric fraction for CRs in a galaxy subject to a starburst-driven outflow. The strongest parameter dependency was found to be on the total galaxy (halo) mass. The outflow was set to have a CR-driving fraction of 50 per cent as a fiducial choice, with the remaining driving power contributed by thermal gas pressure. \vspace{0.5cm}}
\label{fig:calorimetry}}
{\includegraphics[width=8cm]{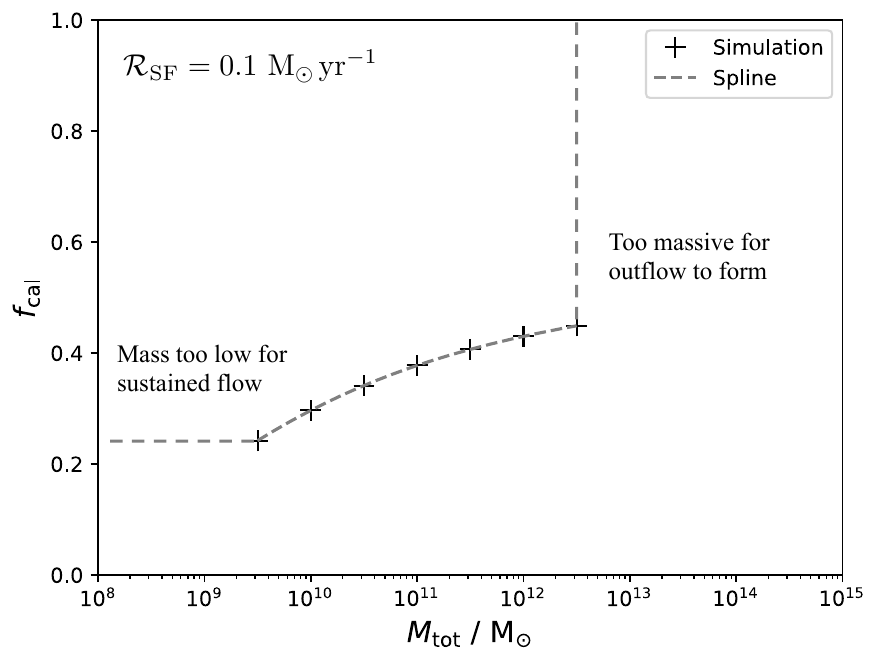}}
\end{figure}

To determine CR calorimetry in our prototype model, we compared the CR energy density at the starburst radius, $r_{\rm sb}$, with that at the centre of the system. We considered that the fractional reduction serves as an appropriate measure 
of the decreased CR calorimetry in a galaxy brought about by the effects of advection in a galactic outflow. 
The effective calorimetric fraction of a galaxy in our model configuration was most strongly affected by total galaxy halo mass (see Fig.~\ref{fig:calorimetry}), with star-formation rate having a less significant effect. In our calculations, we allow for variation of both of these parameters by interpolating over the pre-computed grid of calorimetric fractions. Corresponding $\gamma$-ray luminosities of SFGs were then be adjusted to account for calorimetry in the presence of galactic outflows when modeling the SFG contribution to the EGB.

\section{Results}

\begin{figure*}
    \centering
    \includegraphics[width=\textwidth]{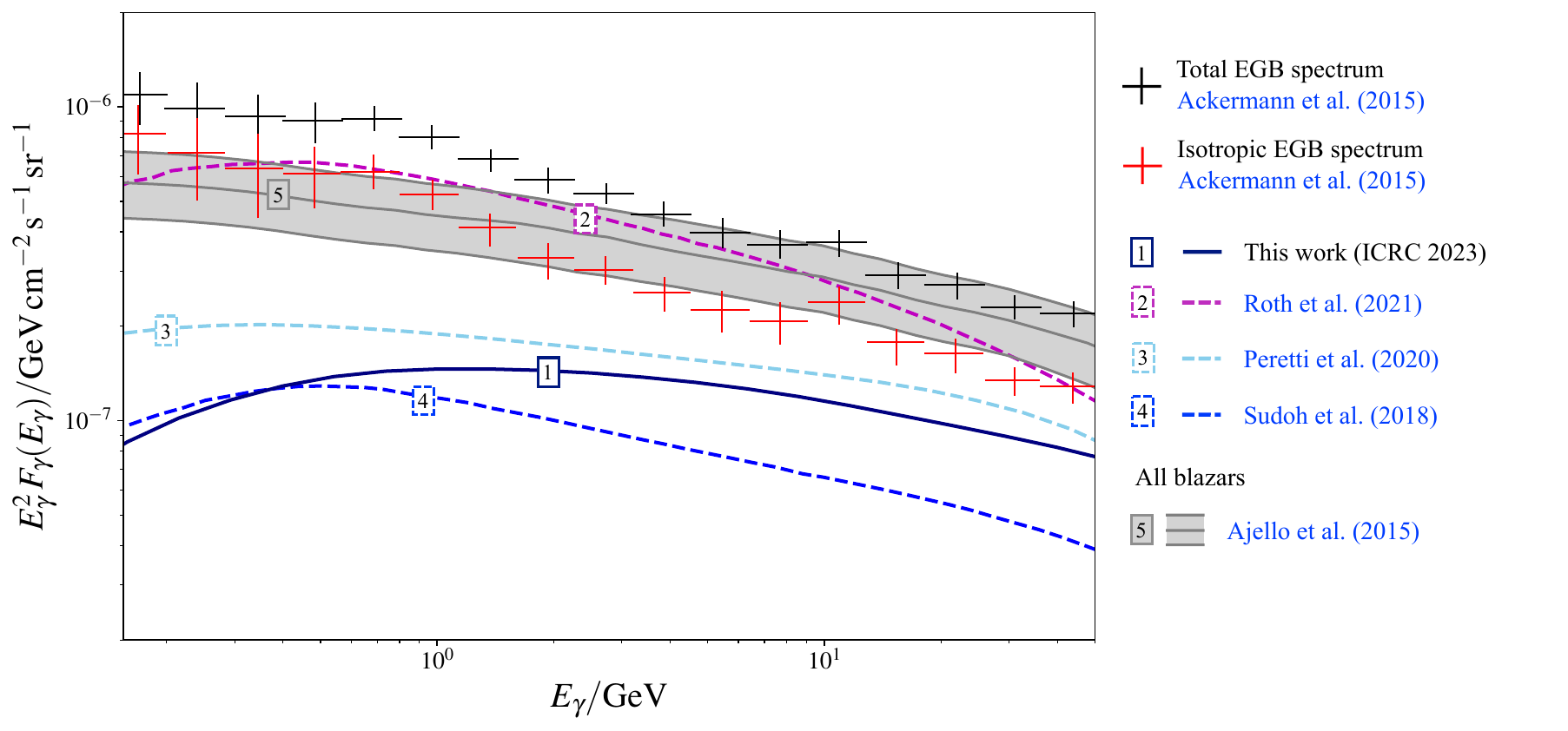}
    \caption{Total contribution from SFGs to the {isotropic} EGB, between 0.1 and 50 GeV. The result from this work is given by line 1, which is in agreement with constraints from the contribution of resolved 
    and unresolved blazars (grey band, denoting the three models of Ref.~\cite{Ajello2015ApJ})
    and the observed EGB with 50 months of \textit{Fermi}-LAT data~\citep{Ackermann2015ApJ}, determined using their foreground model A. Comparison is made with four recent works; Ref.~\cite{Roth2021Natur} (line 2), Ref.~\cite{Peretti2020MNRAS} (line 3), and Ref.~\cite{Sudoh2018PASJ} (line 4).}
    \label{fig:spectrum_compare_appdx}
\end{figure*}

\begin{figure}
    \centering
    \includegraphics[width=0.49\textwidth]{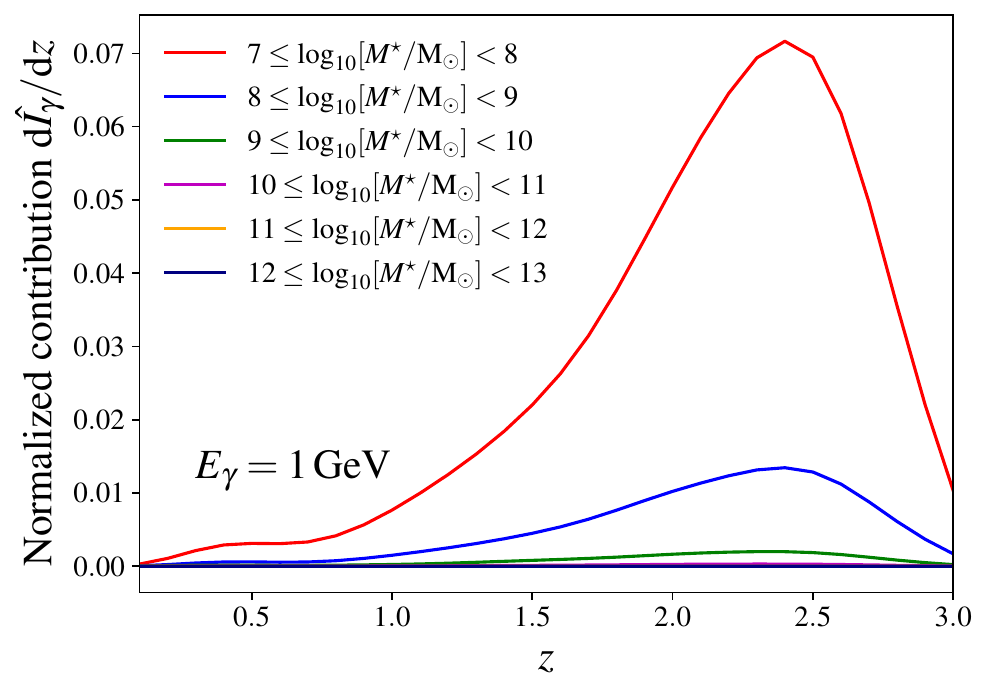}
    \includegraphics[width=0.49\textwidth]{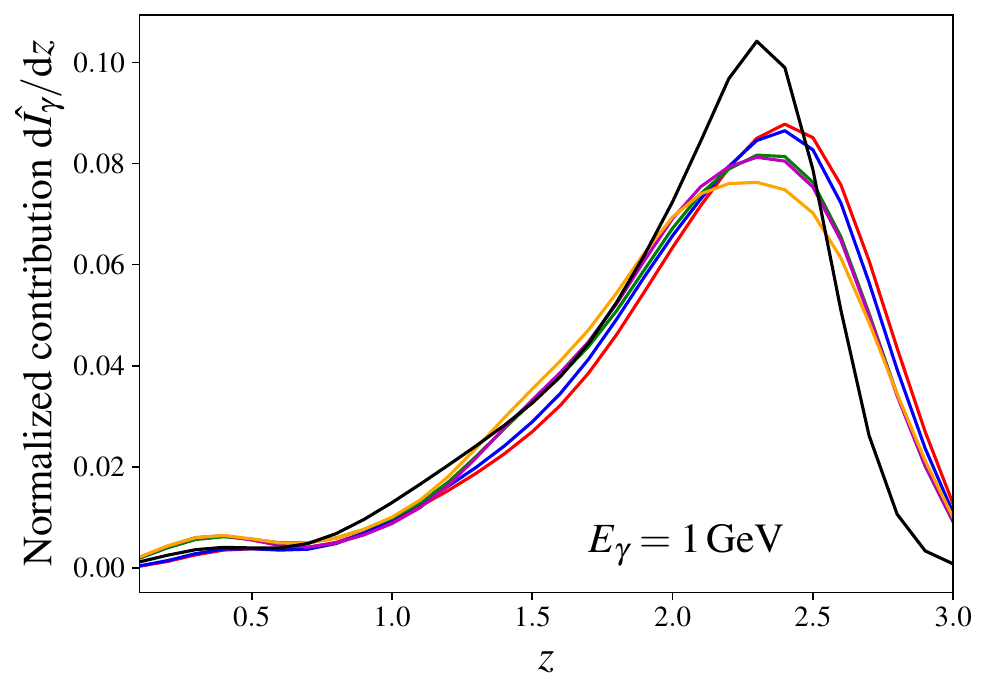}
    \caption{\textbf{Left:} Intensity contributions to the EGB 
    from mass mass-separated SFGs over redshift, normalised to the total EGB intensity from all SFGs at 1 GeV. The 
    majority of the emission originates from low mass galaxies, $M^{\star} = 10^7 - 10^8 \;\!{\rm M}_{\odot}$, reaching a peak at around $z\sim 2.5$. \textbf{Right:} Same as the left panel, but normalized to the SFG $\gamma$-ray intensity in each mass band.}
    \label{fig:normed_1GeV_z}
\end{figure}

\vspace{-0.2cm}
\noindent
Following the same method as Ref.~\cite{Owen2021MNRAS}, we integrate the contribution of populations of SFGs over redshift, from a maximum distance of $z_{\rm max} = 3$. It was shown by Ref.~\cite{Owen2022MNRAS} that the contribution of SFGs to the EGB from distances greater than this is negligible. The resulting EGB spectrum between 0.1 and 50 GeV was computed at $z=0$, and is shown in Fig.~\ref{fig:spectrum_compare_appdx}, in comparison with previous studies and constraints obtained from resolved \textit{Fermi}-LAT sources. 
Fig.~\ref{fig:normed_1GeV_z} 
breaks down the total EGB contribution from SFGs 
according to galaxy mass. This demonstrates that low mass galaxies dominate the SFG contribution, even though they are most capable of sustaining a stable outflow and experience poor CR calorimetry. 
We find the emission from this class of SFGs also can be attributed to a relatively small number of  particularly intensive starbursts. A large majority of the SFG contribution to the EGB is therefore produced by a relatively small number of extreme galaxies in our model.  

\section{Conclusions and discussion}
\label{sec:discussion}

\vspace{-0.2cm}
\noindent
We considered the SFG contribution to the EGB using a prototype modeling approach, with particular 
focus on CR calorimetry, accounting for CR escape from galaxies by diffusion and advection. Compared to previous studies~\cite[in particular,][]{Owen2021MNRAS, Owen2022MNRAS}, this work introduced a more self-consistent treatment of the effects brought about by galactic outflows on CR transport. 
While the assumptions underlying this new model are arguably as arbitrary as assuming a constant CR escape fraction  (as in~\cite{Owen2022MNRAS}), 
our new approach provides a clearer framework 
for building more thorough and self-consistent EGB models. It also highlights areas where improved theoretical understanding is necessary to enhance the reliability of future EGB model advancements, specifically regarding the phase configuration of galactic outflows.

Our study demonstrates that a significant portion of the EGB can be attributed to unresolved populations of SFGs. The majority of this emission originates from low-mass galaxies that exhibit vibrant star-forming activities, occurring just prior to the cosmic `high noon'. These galaxies are relatively rare. This indicates that further consideration is needed to establish how effectively statistical methods can differentiate between contributions from various source populations in the EGB (e.g. SFGs compared to Seyferts and blazars), as the relative dominance of the Poisson term (shot noise) in the EGB angular power spectra due to different source populations may be less distinct than typically assumed. 

\vspace{-0.2cm}
\acknowledgments

\vspace{-0.2cm}
\noindent
ERO is an international research fellow under the Postdoctoral Fellowship of the Japan Society for the Promotion
of Science (JSPS), supported by JSPS KAKENHI Grant Number JP22F22327. This work was achieved in part through the use of large-scale computer systems at the Cybermedia Center, Osaka University, under Research Proposal-based Use (Number 2023-JPC-A02). The software {\tt FLASH} used in this work was developed in part by the DOE NNSA ASC-
and DOE Office of Science ASCR-supported Flash Center for Computational Science at the University of Chicago. 

\bibliographystyle{ICRC}
\bibliography{references}

\providecommand{\href}[2]{#2}\begingroup\raggedright\begin{thebibliography}{10}

\bibitem{Owen2018MNRAS}
E.R.~{Owen}, I.B.~{Jacobsen} et~al., \emph{{Interactions between
  ultra-high-energy particles and protogalactic environments}},
  \href{https://doi.org/10.1093/mnras/sty2279}{\emph{MNRAS} {\bfseries 481}
  (2018) 666} [\href{https://arxiv.org/abs/1808.07837}{{\ttfamily
  1808.07837}}].

\bibitem{Ajello2020ApJ_SFG}
M.~{Ajello}, M.~{Di Mauro} et~al., \emph{{The {\ensuremath{\gamma}}-Ray
  Emission of Star-forming Galaxies}},
  \href{https://doi.org/10.3847/1538-4357/ab86a6}{\emph{ApJ} {\bfseries 894}
  (2020) 88} [\href{https://arxiv.org/abs/2003.05493}{{\ttfamily 2003.05493}}].

\bibitem{Owen2021MNRAS}
E.R.~{Owen}, K.-G.~{Lee} et~al., \emph{{Characterizing the signatures of
  star-forming galaxies in the extragalactic {\ensuremath{\gamma}}-ray
  background}}, \href{https://doi.org/10.1093/mnras/stab1707}{\emph{MNRAS}
  {\bfseries 506} (2021) 52}
  [\href{https://arxiv.org/abs/2106.07308}{{\ttfamily 2106.07308}}].

\bibitem{Owen2022MNRAS}
E.R.~{Owen}, A.K.H.~{Kong} et~al., \emph{{The extragalactic
  {\ensuremath{\gamma}}-ray background: imprints from the physical properties
  and evolution of star-forming galaxy populations}},
  \href{https://doi.org/10.1093/mnras/stac1079}{\emph{\mnras} {\bfseries 513}
  (2022) 2335} [\href{https://arxiv.org/abs/2112.09032}{{\ttfamily
  2112.09032}}].

\bibitem{Roth2021Natur}
M.A.~{Roth}, M.R.~{Krumholz} et~al., \emph{{The diffuse
  {\ensuremath{\gamma}}-ray background is dominated by star-forming galaxies}},
  \href{https://doi.org/10.1038/s41586-021-03802-x}{\emph{\nat} {\bfseries 597}
  (2021) 341} [\href{https://arxiv.org/abs/2109.07598}{{\ttfamily
  2109.07598}}].

\bibitem{Sudoh2018PASJ}
T.~{Sudoh}, T.~{Totani} et~al., \emph{{High-energy gamma-ray and neutrino
  production in star-forming galaxies across cosmic time: Difficulties in
  explaining the IceCube data}},
  \href{https://doi.org/10.1093/pasj/psy039}{\emph{\pasj} {\bfseries 70} (2018)
  49} [\href{https://arxiv.org/abs/1801.09683}{{\ttfamily 1801.09683}}].

\bibitem{Peretti2020MNRAS}
E.~{Peretti}, P.~{Blasi} et~al., \emph{{Contribution of starburst nuclei to the
  diffuse gamma-ray and neutrino flux}},
  \href{https://doi.org/10.1093/mnras/staa698}{\emph{\mnras} {\bfseries 493}
  (2020) 5880} [\href{https://arxiv.org/abs/1911.06163}{{\ttfamily
  1911.06163}}].

\bibitem{Lamastra2017A&A}
A.~{Lamastra}, N.~{Menci} et~al., \emph{{Extragalactic gamma-ray background
  from AGN winds and star-forming galaxies in cosmological galaxy-formation
  models}}, \href{https://doi.org/10.1051/0004-6361/201731452}{\emph{\aap}
  {\bfseries 607} (2017) A18}
  [\href{https://arxiv.org/abs/1709.03497}{{\ttfamily 1709.03497}}].

\bibitem{Ambrosone2021MNRAS}
A.~{Ambrosone}, M.~{Chianese} et~al., \emph{{Starburst galaxies strike back: a
  multi-messenger analysis with Fermi-LAT and IceCube data}},
  \href{https://doi.org/10.1093/mnras/stab659}{\emph{MNRAS} {\bfseries 503}
  (2021) 4032} [\href{https://arxiv.org/abs/2011.02483}{{\ttfamily
  2011.02483}}].

\bibitem{Krumholz2020MNRAS}
M.R.~{Krumholz}, R.M.~{Crocker} et~al., \emph{{Cosmic ray transport in
  starburst galaxies}},
  \href{https://doi.org/10.1093/mnras/staa493}{\emph{\mnras} {\bfseries 493}
  (2020) 2817} [\href{https://arxiv.org/abs/1911.09774}{{\ttfamily
  1911.09774}}].

\bibitem{Melioli2013MNRAS}
C.~{Melioli}, E.M.~{de Gouveia Dal Pino} et~al., \emph{{Evolution of M82-like
  starburst winds revisited: 3D radiative cooling hydrodynamical simulations}},
  \href{https://doi.org/10.1093/mnras/stt126}{\emph{\mnras} {\bfseries 430}
  (2013) 3235} [\href{https://arxiv.org/abs/1301.5005}{{\ttfamily 1301.5005}}].

\bibitem{Martin1998ApJ}
C.L.~{Martin}, \emph{{The Impact of Star Formation on the Interstellar Medium
  in Dwarf Galaxies. II. The Formation of Galactic Winds}},
  \href{https://doi.org/10.1086/306219}{\emph{\apj} {\bfseries 506} (1998) 222}
  [\href{https://arxiv.org/abs/astro-ph/9804165}{{\ttfamily
  astro-ph/9804165}}].

\bibitem{Yu2020MNRAS}
B.P.B.~{Yu}, E.R.~{Owen} et~al., \emph{{A hydrodynamical study of outflows in
  starburst galaxies with different driving mechanisms}},
  \href{https://doi.org/10.1093/mnras/staa021}{\emph{MNRAS} {\bfseries 492}
  (2020) 3179} [\href{https://arxiv.org/abs/2001.04384}{{\ttfamily
  2001.04384}}].

\bibitem{Zaritsky2023MNRAS}
D.~{Zaritsky} and P.~{Behroozi}, \emph{{Photometric mass estimation and the
  stellar mass-halo mass relation for low mass galaxies}},
  \href{https://doi.org/10.1093/mnras/stac3610}{\emph{\mnras} {\bfseries 519}
  (2023) 871} [\href{https://arxiv.org/abs/2212.02948}{{\ttfamily
  2212.02948}}].

\bibitem{Leitherer1999ApJS}
C.~{Leitherer}, D.~{Schaerer} et~al., \emph{{Starburst99: Synthesis Models for
  Galaxies with Active Star Formation}},
  \href{https://doi.org/10.1086/313233}{\emph{\apjs} {\bfseries 123} (1999) 3}
  [\href{https://arxiv.org/abs/astro-ph/9902334}{{\ttfamily
  astro-ph/9902334}}].

\bibitem{Yu2021MNRAS}
B.P.B.~{Yu}, E.R.~{Owen} et~al., \emph{{Outflows from starburst galaxies with
  various driving mechanisms and their X-ray properties}},
  \href{https://doi.org/10.1093/mnras/stab2738}{\emph{\mnras} {\bfseries 508}
  (2021) 5092} [\href{https://arxiv.org/abs/2109.09764}{{\ttfamily
  2109.09764}}].

\bibitem{Fryxell2000ApJS}
B.~{Fryxell}, K.~{Olson} et~al., \emph{{FLASH: An Adaptive Mesh Hydrodynamics
  Code for Modeling Astrophysical Thermonuclear Flashes}},
  \href{https://doi.org/10.1086/317361}{\emph{\apjs} {\bfseries 131} (2000)
  273}.

\bibitem{Ajello2015ApJ}
M.~{Ajello}, D.~{Gasparrini} et~al., \emph{{The Origin of the Extragalactic
  Gamma-Ray Background and Implications for Dark Matter Annihilation}},
  \href{https://doi.org/10.1088/2041-8205/800/2/L27}{\emph{ApJ} {\bfseries 800}
  (2015) L27} [\href{https://arxiv.org/abs/1501.05301}{{\ttfamily
  1501.05301}}].

\bibitem{Ackermann2015ApJ}
M.~{Ackermann}, M.~{Ajello} et~al., \emph{{The Spectrum of Isotropic Diffuse
  Gamma-Ray Emission between 100 MeV and 820 GeV}},
  \href{https://doi.org/10.1088/0004-637X/799/1/86}{\emph{ApJ} {\bfseries 799}
  (2015) 86} [\href{https://arxiv.org/abs/1410.3696}{{\ttfamily 1410.3696}}].

\end{thebibliography}\endgroup

\end{document}